\documentclass{rmaa}
\usepackage{paralist,float}
\usepackage{psfrag,color}
\usepackage[latin1]{inputenc}
\usepackage{amsmath}
\usepackage{mathtools}

\title{3D Modeling of Spectra and Light Curves of Hot Jupiters with PHOENIX; a First Approach}

\author{Juan J. Jim\'enez-Torres,\altaffilmark{1} 
\altaffiltext{1}{Hamburger Sternwarte, Gojenbergsweg 112, 21029 Hamburg, Germany}}

\shortauthor{Jim\'enez-Torres, J. J.}
\shorttitle{Spectra and Light Curves of Hot Jupiters}

\fulladdresses{
\item Juan J. Jim\'enez-Torres: Hamburger Sternwarte, Gojenbergsweg 112, 21029 Hamburg, Germany (jtorres@hs.uni-hamburg.de)}

\listofauthors{J. J. Jim\'enez-Torres}
\indexauthor{Jim\'enez-Torres, J. J.}

\resumen{En este art\'iculo, un modelo de circulaci\'on global fue
  empleado para alimentar el c\'odigo PHOENIX para calcular espectros
  y curvas de luz tridimensionales de Jupiters calientes. Flujos
  radiativos de atm\'osferas con polvo y sin nubes para el planeta
  HD179949b fueron modelados para mostrar diferencias entre ellos. Las
  simulaciones con el c\'odigo PHOENIX pueden explicar las propiedades
  generales de las curvas de luz observadas a 8 $\mu$m, incluyendo el
  hecho de que la raz\'on de flujo planeta-estrella alcance su
  m\'aximo antes del eclipse secundario. El espectro de transmisi\'on
  calculado con PHOENIX reproduce datos observacionales a 3.6 $\mu$m y
  sobreestima valores a 4.5, 5.8 y 8.0 $\mu$m. Estas discrepancias
  provienen de la composici\'on qu\'imica y proporcionan una
  motivaci\'on para incorporar diferentes metalicidades en estudios
  futuros.}

\abstract{In this paper, a detailed Global Circulation Model was
  employed to feed the PHOENIX code to calculate 3D spectra and light
  curves of hot Jupiters. Cloud free and dusty radiative fluxes for
  the planet HD179949b were modeled to show differences between
  them. The PHOENIX simulations can explain the broad features of the
  observed 8 $\mu$m light curves, including the fact that the
  planet-star flux ratio peaks before the secondary eclipse. The
  PHOENIX reflection spectrum matches the Spitzer secondary-eclipse
  depth at 3.6 $\mu$m and underpredicts the eclipse depths at 4.5, 5.8
  and 8.0 $\mu$m. These discrepancies result from the chemical
  composition and provide motivation for incorporating different
  metallicities in future studies.}

\addkeyword{Planets and satellites: atmospheres -- Planets and
  satellites: individual (HD179949b, HD209458b)}

\begin{document}

\maketitle

\section{Introduction}\label{sectionintroduction}

A hot Jupiter is any of various extrasolar planets which is as big and
gaseous as Jupiter, but is much hotter due to a very close orbit
around its parent star (Seager 2010). Hot jupiter planets experience a
strong irradiation from their parent stars and are expected to be
tidally locked because they are very close ($<$ 0.05 AU) to their host
star, resulting in permanent day and nightsides (Knutson et
al. 2007). The consequences of such small orbital separations are
atmospheric structures and spectra that are very different from the
giant planets in the Solar system (Barman et al. 2005).

In the literature, there are several models of spectra and light
curves of hot Jupiters. The majority of these models are one
dimensional (1D) codes, e.g., Burrows et al. (2006) investigated
spectra and light curves of planets with various day-night effective
temperature differences, assuming 1D profiles for each hemisphere. Iro
et al. (2005) extended 1D models by adding heat transport due to a
simple parameterization of winds to generate longitude-dependent
temperature maps, but they did not compute disk-averaged spectra for
these models. Barman et al.(2005) investigated two-dimensional (2D)
models with axial symmetry around the planet's substellar-antistellar
axis and computed infrared spectra as a function of orbital
phase. These 1D and 2D radiative-convective equilibrium models have
had some success in matching Spitzer observations; however, Seager et
al. (2005) and Deming et al. (2006) showed that ground-based data for
HD209458b do not indicate prominent flux peaks at 2.3 and 3.8 $\mu$m,
which the 1D and 2D solar composition models predicted. Dobbs-Dixon
$\&$ Agol (2013) presented 3D radiative-hydrodynamical models for the
hot Jupiter HD189733b. To address the radiative transfer they
developed a frequency dependent, two-stream approximation (Mihalas
1978) for the radial radiative flux. Their transit spectrum agrees
well with the data from the infrared to the UV, though it slightly
under-predicts the observations at wavelengths shorter than $\sim$ 0.6
$\mu$m. Their emission spectrum agrees well at 5.8 and 8 $\mu$m, but
over-predicts the emission at 3.6 and 4.5 $\mu$m. Their phase curves
agree fairly well with the amplitudes of variations, shape, and phases
of minimum and maximum flux. However, they over-predict the peak
amplitude at 3.6 $\mu$m and 4.5 $\mu$m. Rauscher $\&$ Menou (2010)
presented a 3D hot Jupiter model, extending from 200 bar to 1 mbar,
using an intermediate general circulation model and adopting a
physical setup nearly identical to the model of HD 209458b by Cooper
$\&$ Showman (2006). Discrepancies can have their origin in the
different methods of solution adopted and/or the fact the equations
solved are missing some of the physics at work in the hot Jupiter
context. In the hot Jupiter modeling context, Rauscher $\&$ Menou can
not be certain of the accuracy and validity of their models until they
can likewise produce consistent results. By using a 3D general
circulation model with Newtonian cooling and dayside and nightside
equilibrium temperature profiles, Burrows et al. (2010) created models
for the planet HD209458b. They obtained transit radius spectra during
the primary transit. With a 1D spectral atmosphere code, they
integrated over the face of the planet seen by an observer at various
orbital phases and calculated light curves as a function of
wavelength. However, since their circulation model uses Newtonian
cooling (and not radiative transfer using opacities that correspond to
those used in the post-processing), their calculations could be
slightly inconsistent. On their behalf, for the planets HD209458b and
HD189733b, Fortney et al. (2010) computed spectra for both 1D and 3D
model atmospheres and examined the differences between them.

In this paper, the 3D version of the radiative transfer PHOENIX code
has been used to compute light curves as well as reflection spectra
for a 3D temperature-pressure structure on a latitude and longitude
grid. A suitable treatment of atmospheric dynamics, non-equilibrium
chemistry, and 3D radiative transfer is a considerable improvement in
this work. In this scheme, given the 3D Tempertaure-Pressure grid,
with corresponding chemical mixing ratios, the PHOENIX radiative
transfer scheme solves for the upward and downward fluxes in each
layer. These fluxes are wavelength dependent and differ from layer to
layer. The modeling of atmospheres with the PHOENIX code can be
important for at least two reasons. First, since the emergent spectra
of hot Jupiters are determined by the chemistry and physics of their
outer atmospheres, when direct detection of hot Jupiters is achieved
and spectra are obtained, PHOENIX models can be essential in the
interpretation of the data and in the extraction of essential physical
information such as radius, gravity, temperature, and
composition. Second, theoretical spectral models are important in
guiding current and upcoming direct hot Jupiter searches.

This work is organized in 3 sections: the first one provides a brief
introduction to hot Jupiters and the relevance of the PHOENIX code to
model planetary atmospheres. The second section provides the
description of the PHOENIX code to simulate the irradiation of hot
Jupiters, explains the mathematical framework to solve the radiative
transfer equation by numerical methods, and also gives a description
of the physics and microphysics of irradiated atmospheres. The Global
Circulation Model (GCM) to generate the temperature-pressure
structures to feed the 3D radiative transfer PHOENIX code is also
presented in section 2. Section 3 corresponds to results; it shows
radiative fluxes at interesting positions on the planet; cloud-free
and dusty atmosphere models of the hot Jupiter HD179949b are presented
to see the differences between them. PHOENIX cloud-free and dusty
light curves were calculated in different wavelength bands and
compared to actual data points. Finally, cloud-free and dusty spectra
are presented for the planet HD179949b. The cloud-free PHOENIX
spectrum was confronted with observational data from the hot jupiter
HD209242b and it was found that the theoretical model is able to fit
some planet-star flux data points.

\section{Methodology}\label{sectionmethodology}

For calculations in this paper, the 3D radiative transfer equation is
solved along rays or characteristics (see Hauschildt $\&$ Baron 2006;
Olson $\&$ Kunasz 1987). The iterative method for the solution of the
radiative transfer equation is the method to solve the scattering
problem and is based on the philosophy of operator perturbation
(Cannon 1973; Scharmer 1984). Its numerical approach is based on
finding consistent solutions of the source function and the mean
intensity by iterative methods. The mean intensities in this study
were calculated as established in Hauschildt $\&$ Baron (2006), that
is, the mean intensity $J$ is obtained from the source function $S$ by
a formal solution of the radiative transfer equation. The equation of
state (EOS) calculations themselves follow the method discussed in
Smith $\&$ Missen (1982), whereas the Equation of State (EOS) plus
opacity setups are identical to Husser et al. (2013).

\subsection{The PHOENIX code}\label{subsectionphoenixcode}

The latest version of the PHOENIX code 10.6 has been used to calulate
the mean intensities, reflection spectra and light curves of hot
Jupiters. The original version of PHOENIX includes a detailed
radiative transfer (Hauschildt 1992) that allows for spherical
symmetry. The modeling of the effects of extremely irradiated
atmospheres considers a full 3D radiation transport modeling,
including deviations from the local thermodynamics equilibrium
(NLTE). The treatment includes scattering, which is an important
effect for the day-side of a planet due to the high temperature caused
by the irradiation and near the terminator due to scattering on dust
particles that can survive there due to the lower termperature. The
numerical atmosphere modeling with PHOENIX includes the combination of
a radiative transfer technique with atmospheric structure and boundary
constraints. The PHOENIX code incorporates a 3D radiation transport
framework (Hauschildt $\&$ Baron 2006; Baron $\&$ Hauschildt 2007;
Hauschildt et al. 2008; Baron et al. 2009) into the PHOENIX package,
creating the 3D PHOENIX mode in addition to the 1D PHOENIX mode of the
PHOENIX program package (Hauschildt $\&$ Baron 2010).

Planetary atmospheres in this work are modeled in spherical
coordinates whose volume is divided into discrete elements of
space. Such a finite element of space is called a volume element or
voxel. A single characteristic hits the voxels of the grid under
different angles and thus, provides different contributions to the
angular resolution of the mean intensity in the different layers. Each
voxel is then assigned with physical quantities like temperature,
pressure, opacity, emissivity, intensity, source function and so on,
then the radiative transfer is solved inside each voxel by a numerical
quadrature (see Barman et al. 2005).

\subsection{The Global Circulation Model}\label{gcmsubsection}

The impact of horizontal motions on hot Jupiter atmospheres has been
modeled by a variety of groups (e.g. Showman \& Guillot 2002; Cho et
al. 2003; Burkert et al. 2005; Cooper \& Showman 2005); despite the
differences in methods and results, the general consensus from these
hydrodynamic simulations is that circulations can redistribute a
fraction of the incident energy over large portions of a strongly
irradiated planetary atmosphere.

For models in this paper, the 3D approaches to the atmospheric
circulation of hot Jupiters are external models (T. Barman, 2012,
private communication), that is, the Global Circulation Models in this
work were generated from the model data by Showman et al. (2009), just
straight from them. They approximated the dayside heating and
nightside cooling using a Newtonian cooling/heating scheme, which
parameterizes the radiative heating rate (in K s$^{-1}$) as (T$_{eq}$
- T)/$\tau_{rad}$, where T$_{eq}$ is the specified
radiative-equilibrium temperature profile, $T$ is the actual
temperature, and $\tau_{rad}$ is the radiative-equilibrium
timescale. These atmospheric circulation models are based on six
fundamental conservation equations: the conservation of mass,
conservation of momentum (one equation for each dimension),
conservation of energy, and the ideal gas law as the equation of
state. Within the context of the Newtonian cooling framework, Showman
et al. (2009) calculated the longitude, latitude, and height-dependent
radiative-equilibrium temperatures.

\begin{figure}[H]
\centering
\includegraphics[height=14cm]{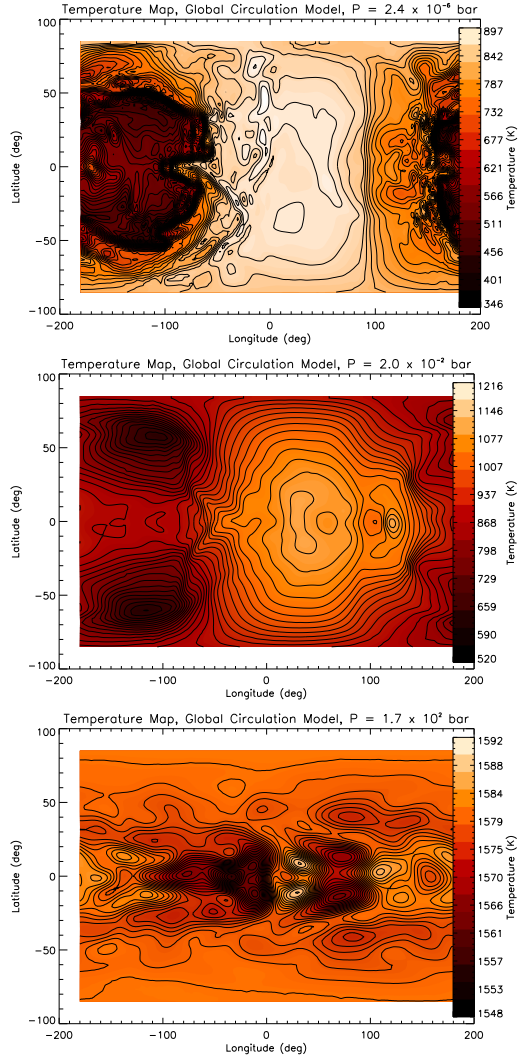}
\vspace{0.1cm}
\caption{\scriptsize Simulated temperature maps of the Global
  Circulation Model. Maps use a linear scale, with the hottest points
  in white and the coldest points in black. From top to bottom there
  are three isobars: 0.0024, 20 and 170 000 mbar. The substellar point
  is at (0,0) in longitude and latitude. The dynamics distorts the
  temperature pattern in a complex manner; this generally includes an
  eastward displacement of the hottest regions from the substellar
  point.}
\label{fig.Temp_Map}
\end{figure}

The radiative-equilibrium temperature structures described above were
used as input structure files to feed the 3D radiative transfer
PHOENIX code to generate models of hot Jupiters. Figure
\ref{fig.Temp_Map} illustrates the behavior for the simulation of the
Showman et al. (2009) Global Circulation Model using solar
opacities. It shows the temperature (colorscale) over the globe on
three different pressure levels. A temperature range of 346 K to 896 K
is observed on the external layer with pressure equal to $\sim$
10$^{-6}$ bar, indicating that the absorbed energy from the parent
star is distributed through the planet's atmosphere, that is, the
irradiated dayside is efficiently redistributed throughout the
atmosphere. The day-night temperature differences reached nearly 40 K
at 170 bar and 600 K at 20 mbar. Although the detailed structure
varies between levels, the hottest regions lie east of the substellar
point throughout. The atmospheric dynamics distorts the temperature
pattern in a complex manner, which includes the eastward displacement
of the hottest regions from the substellar point.

\begin{figure}[H]
\centering
\includegraphics[width=12cm]{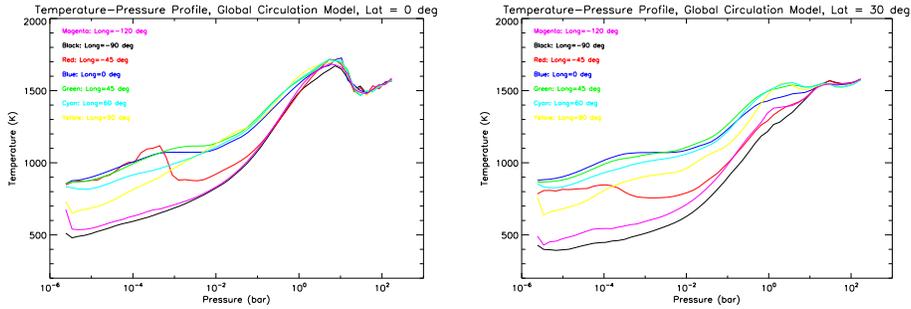}
\vspace{0.1cm}
\caption{\scriptsize A selection of temperature-pressure profiles for
  the solar-abundance Global Circulation Model. The equatorial region
  is on average warmer than the 30 deg latitude at pressures of less
  than a few bars. Note the formation of a stratosphere at pressures
  less than 0.1 bars.}
\label{fig.Temp_Press_Opaci}
\end{figure}

Figure \ref{fig.Temp_Press_Opaci}, illustrates a diversity of vertical
temperature profiles that occur. It portrays in two panels the
temperature-pressure structure profile as function of longitude. The
left frame corresponds to latitude 0 degrees and the right one to 30
degrees. Solid lines correspond to different positions on the planet;
magenta line corresponds to Long=-120 deg, black to Long=-90 deg, red
to Long=-45 deg, blue to Long=0 deg, green to Long=45 deg, cyan to
Long=60 deg and yellow to Long=90 deg. Slight bumps (on the nightside)
and depressions (on the dayside) are noted between $\sim$ 10$^1$ and
$\sim$ 10$^2$ bars in the temperature-pressure profiles, this region
is near where the heat redistribution is expected to occur.

As expected, the equatorial region is on average warmer than the 30
deg latitude at pressures of less than a few bars. At low pressures,
longitudinal variation is comparable to the latitudinal variation. The
temperature declines smoothly with altitude from $\sim$ 10$^{0}$ bars
to $\sim$ 10 mbar. This has important implications for the spectra and
light curves, which originate within this layer. Dynamics modifies the
deep stable radiative layer from 10-100 bars, leading to significant
latitude variation of both temperature and static stability (Showman
et al. 2009). Following analysis found in Seager (2010), the
Temperature-Pressure profiles shown in figure
\ref{fig.Temp_Press_Opaci} could be divided into three representative
layers. Above layer 1, at pressures below P $\sim$ 10$^{-2}$, the
optical depth at all wavelengths becomes low enough so that the layers
of the atmosphere are transparent to the incoming and outgoing
radiation and not relevant for spectral features. This uppermost
layer, has no thermal inversions. The layer 2 from $\sim$10$^{-2}$ bar
to $\sim$10$^0$ bar is where most spectral features are formed. In
this layer, the temperature structure is governed by radiative process
and possibly by atmospheric dynamics. These optically thin layers are
at altitudes where thermal inversions may be formed. The layer 3 from
$\sim$10$^0$ bar to $\sim$10$^2$ bar is the regime where a high
optical depth leads to radiative diffusion and the related isothermal
temperature structure. Below this layer, in the deepest layers of the
planet atmosphere, convection is the dominant energy transport
mechanism.

\section{Results}\label{sectionresults}

In this section theoretical radiative fluxes, light curves and
reflection spectra are presented to simulate irradiated Global
Circulation Models (hot Jupiters). To calculate the mean intensities,
the temperature-pressure profiles described in subsection
\ref{gcmsubsection} are used as an input for the 3D radiative transfer
PHOENIX code so that it can generate the radiation fields. Then this
grid with $S$ (source function) and $J$ (mean intensities) is used to
create 3D visualizations (spectro-images), spectra and light
curves. In this study, the Global Circulation Model passing in front
of its parent star is simulated with PHOENIX on a 3.5M voxels grid
with $n_r$ = 53, $n_{\theta}$ = 257 and $n_{\phi}$ = 257 voxels. This
number of voxels was selected to be large enough to minimize the
effects of pixellation but small enough to give reasonable computation
times. The irradiated models presented in this section were computed
using the PHOENIX atmosphere code (Hauschildt $\&$ Baron 1999; Allard
et al. 2001) adapted to include extrinsic radiation as described in
Barman et al. (2001; 2002).

\subsection{Test radiative fluxes}\label{subsectiontestreflectancespectra}

Experiments in this subsection corresponds to radiative fluxes of the
hot Jupiter HD179949b for which data from the Spitzer and Hubble Space
Telescopes have been published (e.g., Burrows et al. 2008; Cowan et
al. 2007). Physical and orbital parameters in Table
\ref{table.comp_atm_jupiter} were employed to simulate the hot
Jupiter. The parent star HD179949 is a F8V star, orbited by HD179949b
at 0.0443 $\pm$ 0.0026 AU placing it among the closest orbiting
planets in a low-eccentricity (e=0.022$\pm$0.015) orbit. The
scattering parameter in the PHOENIX models was set to $\epsilon$ =
10$^{-4}$ which could be a reasonable value to simulate a strong
scattering environment. To get the radiative fluxes, it is assumed
that, for irradiated atmospheres, the parent star is located at the
zenith over the substellar point, that is, $\theta$ = $\pi$/2 and
$\phi$ = $\pi$, where $\theta$ corresponds to the polar angle and
$\phi$ to the azimuth angle. In this case, the day-night boundary is
at $\theta$ = 0. That is, the parent star is above the substellar
point, 'above' means in the vertical direction assuming no axial
tilts. The PHOENIX models assume circular orbits, which is a good
approximation for most exoplanets in tight orbits.

\begin{center}
\begin{table*}\scriptsize
\caption{\footnotesize Physical and orbital parameters of the
  Planetary System HD179949 (Cowan et al. (2007); Burrows et
  al. 2008)).}
\begin{tabular}{ | p{0.13\linewidth} |
                   p{0.075\linewidth} |
                   p{0.13\linewidth} |
                   p{0.093\linewidth} |
                   p{0.093\linewidth} |
                   p{0.093\linewidth} |
                   p{0.093\linewidth} | }
\hline
Planet&a(AU)&Star&R$_{\star}$(R$_{\odot}$)&Teff(K)&M$_{\star}$(M$_{\odot}$)&L$_{\star}$(L$_{\odot}$)\\
\hline
HD 179949b&0.045&HD 179949&1.24&6200&1.24&2.0\\
\hline
\end{tabular}
\label{table.comp_atm_jupiter}
\end{table*}
\end{center}

The first experiment in this subsection corresponds to 3D cloud free
radiative fluxes for the hot Jupiter HD179949b. The irradiated planets
in this study have been modeled for two extreme cases: one where dust
clouds form and remain suspended in the atmosphere (dusty models), and
another where dust clouds form but completely settle out of the
atmosphere (cloud free models). In the cloud-free situations, dust
grains form in the atmosphere at locations determined by the chemical
equilibrium equations but their opacity contribution is ignored,
mimicking a complete removal of the grains by efficient gravitational
settling. Therefore, those models represent cloud free atmospheres. In
thay way, the cloud-free models may be thought of as clear skies, and
the dusty model as cloudy skies.

\begin{figure}[H]
   \begin{center}
\includegraphics[width=12.2cm]{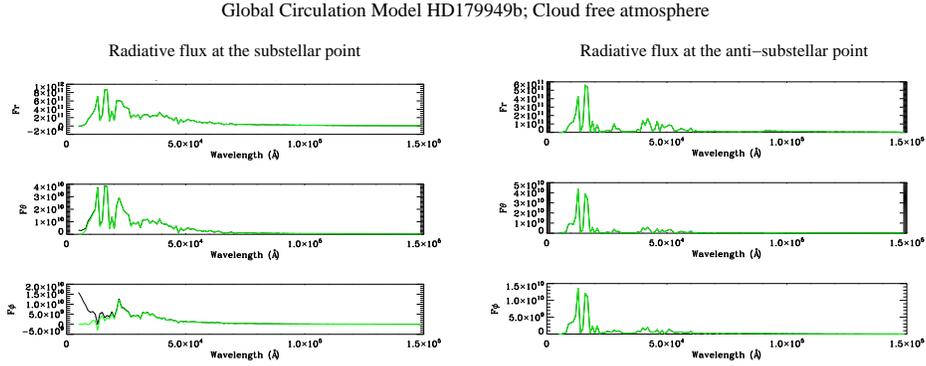} 
  \end{center}
\caption{\scriptsize 3D cloud-free radiative fluxes of the hot Jupiter
  HD179949b in the 5000 - 150000 {\AA} wavelength range. Green curves
  correspond to the 1DRT
Two-Stream model and black curve to the 3DRT PHOENIX model. Irradiated atmospheres on the day side remain very
  transparent and allow large amounts of flux to emerge from the deep
  hotter layers. Nightside spectra at the antisubstellar point show
  deep absorption bands of H$_2$O, CO, and/or CH$_4$; the 1DRT
Two-Stream model
  dominates the nightside regime.}
\label{fig.spectradifferentpositionsnodustcases}
\end{figure}

Green curves in Figure \ref{fig.spectradifferentpositionsnodustcases}
correspond to the Global Circulation Model, here also called the 1DRT
Two-Stream model and black curves to the irradiated planet HD179949b
modeled with PHOENIX, also called the 3DRT PHOENIX model. Plots show
the radial and angular components of the flux vector (F$_r$,
F$_{\theta}$ and F$_{\phi}$) versus wavelength. Figure
\ref{fig.spectradifferentpositionsnodustcases} shows radiative fluxes
at the substellar and antisubstellar points. That is, it does not show
flux vectors across all outermost voxels (outermost radius, all
($\theta$, $\phi$) points). In the outermost voxels the pressure is
equal to 0.002 mbar; temperature is equal to 880 K at the substellar
point and 300 K at the antisubstellar point. Figure
\ref{fig.spectradifferentpositionsnodustcases} shows cloud free fluxes
in the 5000 - 150000 {\AA} wavelength range, left frame corresponds to
the substellar point and right frame to the antisubstellar point;
these cloud free models do include TiO and VO opacities. The 3D
radiative fluxes are medium resolution calculations, covering a broad
wavelength range and placing the overall features in context, while
models at high resolution are needed to identify specific absorption
features.

According to Figure \ref{fig.spectradifferentpositionsnodustcases},
models on the day side show that in the 5000 - 15000 {\AA} wavelength
range the irradiated flux is larger than the emission by the 1DRT
Two-Stream model, this is noted in the $F_{\theta}$ and $F_{\phi}$
flux components. Significant modulations present in the near-IR
spectrum are best explained by the presence of water vapor and the
carbon bearing molecules. The 3DRT PHOENIX model reflects a
considerable amount of radiation in the optical bands compared to the
1DRT Two-Stream model. In models on the night side molecules
predominate in the fluxes. At the anti-substellar point absorption
features are created by molecules that could survive lower
temperatures. The PHOENIX radiative fluxes in the near-IR contain
significant modulations, which can be attributed to the presence of
molecular bands seen in absorption. These are water (H$_2$O), carbon
monoxide (CO), and carbon dioxide (CO$_2$) bands.

The H$_2$O and CH$_4$ molecules have already been inferred for some
hot Jupiters via transmission photometry and spectroscopy (e.g. Desert
et al. 2009, Deming et al. 2013, Gibson et al. 2012, Sing et al. 2009,
Birkby et al. 2013). Recently Madhusudhan et al. (2014) report
conclusive measurements of H$_2$O in the hot jupiter HD209458b, indeed
they report the most precise H$_2$O measurement in an exoplanet to
date that suggests a $\sim$20-135$\times$sub-solar H$_2$O abundance;
and CO, thermochemically the most stable carbon molecule on the hot
day-side, has been inferred from photometry (Barmam 2008; Charbonneau
et al.  2008). The light reflected in the optical regime could be due
to Rayleigh scattering by the two most abundant species, H$_2$ and He
(Lecavelier des Etangs et al. 2008).

By modeling radiative fluxes at different positions on the planet, it
is possible to see that the night side is in most cases equal to the
same voxels in the 1DRT Two-Stream model, with the exception of the
places where the light from the star shines through. Voxels that are
identical (same fluxes) are not directly affected by the light from
the star, the others are directly affected (and carry information
about the atmosphere to the observers when the planet is transiting
the stellar disk). Figure
\ref{fig.spectradifferentpositionsnodustcases} illustrates how
molecular opacities dominate the spectral range through which the
planet radiates the bulk of their emergent flux ($\sim$ 15000 {\AA})
leaving practically no window of true continuum. Models show molecular
absorption features in the 10000 - 30000 {\AA} wavelength
range. Irradiated atmospheres on the day side remain very transparent
and allow large amounts of flux to emerge from the deep hotter
layers. Nightside fluxes at the antisubstellar point show deep
absorption bands of H$_2$O, CO, and/or CH$_4$. In chemical
equilibrium, CO would dominate over CH$_4$ across much of the dayside,
but CH$_4$ dominates on the nightside. Molecular absorptions,
primarily due to water and methane, are visible for $>$ 10000
{\AA}. That is, the methane absorption probes the deeper layers and
the water bands probe the upper layers and are sensitive to the
irradiation.

\begin{figure}[H]
   \begin{center}
\includegraphics[width=12.2cm]{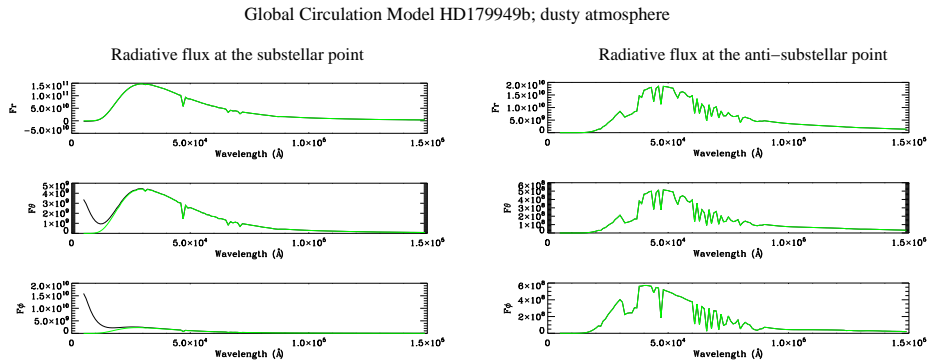} 
  \end{center}
\caption{\scriptsize Same description as Figure
  \ref{fig.spectradifferentpositionsnodustcases} but for models with
  dust opacity. Dust opacity generally produces a hotter atmosphere at
  all depths with a smoother spectral energy distribution. Absorption
  features are more prominent on the night side than on the day
  side. On the day side, the strong heating effects of dust opacities
  prevent the formation of methane bands, and H$_2$O is dissociated
  while producing a hotter water vapor opacity profile, which is much
  weaker and more transparent to radiation.}
\label{fig.spectradifferentpositionsnowithdust}
\end{figure}

The second test experiment in this subsection shows the effect by the
presence of dust in the atmosphere of HD179949b. These cloudy models
refer to the case when dust forms in the atmosphere at locations
determined by chemical equilibrium equations. It is known that the
forming dust clouds provide an enormous opacity and that condensating
solid particles as clouds or haze layers can have strong opacities
(e.g. Jones $\&$ Tsuji 1997). Because of this, it is expected that the
spectral appearance of the object will be considerably changed,
because clouds effectively reflect impinging radiation. The dusty
models with PHOENIX exhibit a complex mixture of cloud species
throughout the atmosphere. According to Barman et al. (2001), Fe,
Mg$_2$SiO$_4$ and MgSiO$_3$ are the most prominent species except for
the deeper layers where CaMgSi$_2$O$_6$ begins to
dominate. Specifically, he argues that in the dusty models, Fe,
Mg$_2$SiO$_4$, MgSiO$_3$, CaMgSi$_2$O$_6$ and MgAl$_2$O$_4$ are
dominant dust species which form a cloudy region extending from
roughly $\tau_{std}$ = 1.0 to the top of the atmosphere, below this
region ($\tau_{std}$ $>$ 1.0) is a mixture of various other
condensates.  To include dust grains, I have assumed an interstellar
size distribution with diameters ranging from 0.00625 to 0.24 microns
and the chemical equilibrium equations incorporate over 1000 liquids
and crystals, details in Barman et al. (2001) and Allard et
al. (2001). Within the dusty calculations, the number densities for
each grain species were calculated using the method of Grossman (1972)
and the Gibbs free energies of formation from the JANAF database
(Chase et al., 1985). Spherical shapes of the grains in this work are
presented for simplicity. The effects of grain sizes and dust porosity
on models are not explored in this study. Figure
\ref{fig.spectradifferentpositionsnowithdust} presents dusty models at
two positions on the planet; left frames correspond to fluxes at the
substellar point and the right frames to fluxes at the antisubstellar
point.

On the day side, the strong heating effects of dust opacities prevent
the formation of methane bands, and H$_2$O is dissociated while
producing a hotter water vapor opacity profile, which is much weaker
and more transparent to radiation. The dusty absorption features are
more prominent at the antisubstellar point than at the stellar
point. The dusty model generates a very smooth flux in the optical, as
expected, for an atmosphere dominated by grain opacity. In general,
cloud free models are much brighter than the dusty cases because the
dust opacity blocks most of the thermal radiation. In dayside models,
for wavelengths shorter than $\sim$ 2000 {\AA}, the reflected light in
the dusty model matches closely the stellar light. The reflected light
in the dusty models could be due almost entirely to Mie scattering
which is a fairly grey process.

\begin{figure}[H]
   \begin{center}
\includegraphics[width=12.2cm]{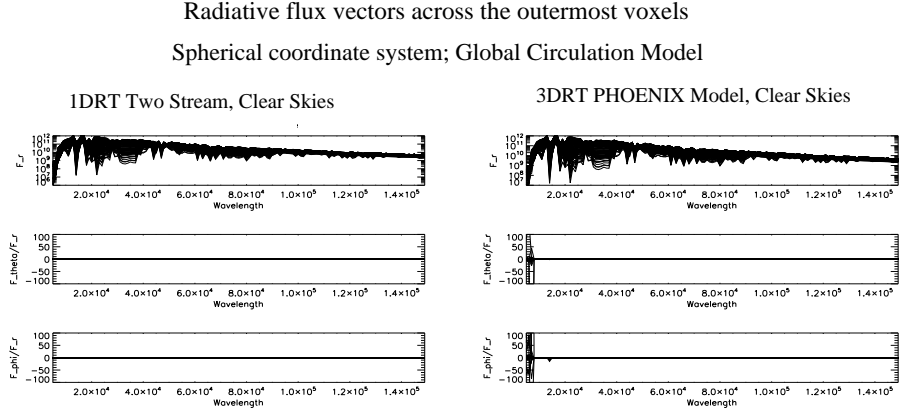} 
  \end{center}
\caption{\scriptsize Radiative flux vectors across the outermost voxels for the 1DRT Two-Stream (left) and 3DRT PHOENIX (right) Models with clear skies. Models are shown in spherical coordinate systems. Calculations use a 3D spherical coordinate system with $n_r$ = 53, $n_{\theta}$ = 257 and $n_{\phi}$ = 257 voxels for a total of about 3.5M voxels. The $F_r$ panels show the radial component of all outer voxels in logarithmic scale.
The bottom panels show the corresponding runs of $F_{\theta}$/$F_r$ and $F_{\phi}$/$F_r$, respectively. They should be identically zero and the deviations measure the internal accuracy. The wavelengths are given in angstroms and
the fluxes are in cgs units.}
\label{fig.collage}
\end{figure}

The radiative fluxes plotted in Figures
\ref{fig.spectradifferentpositionsnodustcases} and
\ref{fig.spectradifferentpositionsnowithdust} show flux vectors at one
specific outermost voxel, that is, they do not show the flux vectors
of all outermost voxels (outermost radius, all ($\theta$, $\phi$)
points). Figure \ref{fig.collage} shows radiative fluxes across the
outermost voxels for the 1DRT Two-Stream (left) and 3DRT PHOENIX
(right) models with clear skies. Models are shown in spherical
coordinate systems. The radial component of the flux vector $F_r$ is
the only non-zero component as in spherical symmetry both $F_{\theta}$
and $F_{\phi}$ are zero. The error in the angular components at short
wavelengths of the 3DRT PHOENIX calculations can be due to the number
of solid angle points. Hauschildt $\&$ Baron (2006, 2010) affirm that
if the number of solid angle points with PHOENIX is too small, then
$F_{\theta}$ and $F_{\phi}$ could have errors. They found that the
higher solid angle resolution reduces the errors in $F_{\theta}$ and
$F_{\phi}$ considerably and the higher internal accuracy due to more
solid angle points also increases the internal accuracy of $F_r$.

\begin{figure}[H]
   \begin{center}
\includegraphics[width=12.25cm]{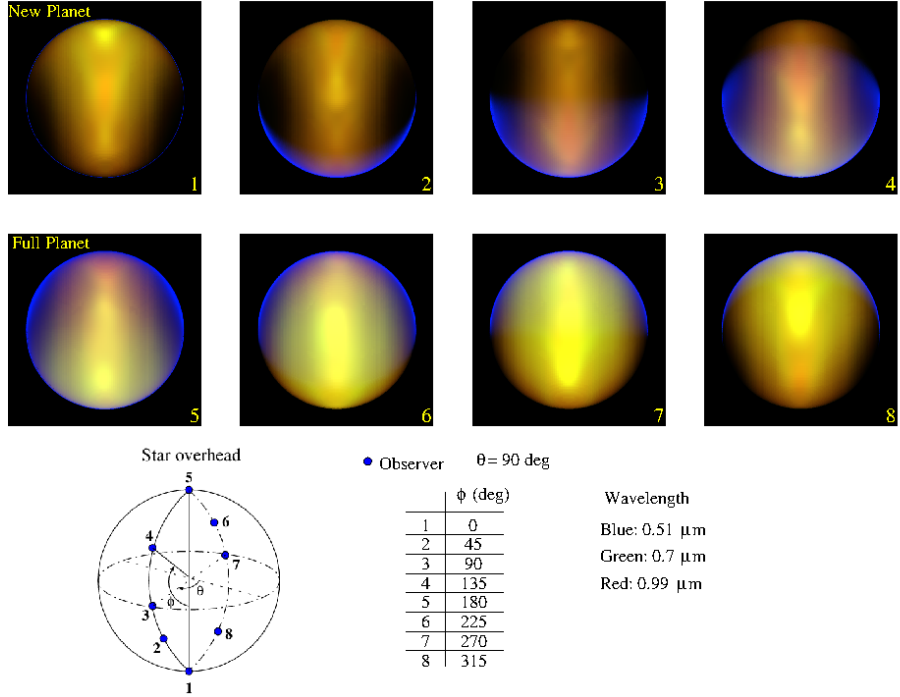} 
  \end{center}
\caption{\scriptsize Visualizations of the results for
  multi-wavelength 3D radiation transfer for a global circulation
  planetary atmosphere model for eight 45-degree phase increments in
  the planet HD179949b. The red channel corresponds to 0.99 $\mu$m,
  the green channel to 0.7 $\mu$m and the blue channel to 0.51
  $\mu$m. The irradiated 1DRT
Two-Stream model predicts a day side that is very similar
  to a blackbody, leading to a whitish appearance. On the night side,
  which is fairly cool, strong methane absorption knocks out the blue
  and green, leaving only red. Spectro-images directly map the patchy
  global patterns inferred to exist from models of multiwavelength
  variability.}
\label{fig.spectroimagesrealplanet}
\end{figure}

The third experiment in this subsection shows a 3D visualization of
the irradiated Global Circulation Model to view spectro-images of the
emitted mean intensities. The visualization reads the results of the
3D radiative transfer, perform a formal solution for a specific
($\theta$,$\phi$) and displays the results as images of the
intensities. The 3D visualizations draw on the Global Circulation
Model described in subsection \ref{gcmsubsection} and show the
influence of Atmospheric Dynamics on infrared radiative fluxes. Figure
\ref{fig.spectroimagesrealplanet} shows wavelength resolved images of
HD179949b for eight 45-degree phase increments. Frames show mean
intensities in the 0.51-0.99 $\mu$m wavelength range (near infrared
band). These kinds of visualizations could be a good step forward for
the computational imaging of extrasolar planets. The slight
pixellation and asymmetries are due to spatial and angular resolution.

In Figure \ref{fig.spectroimagesrealplanet} a bright near-sub stellar
region can clearly be seen in the New Planet panel, the lightest and
darkest regions shown correspond to brightness variations. A natural
explanation for the features seen in Figure
\ref{fig.spectroimagesrealplanet} is that spectro-images are directly
mapping the patchy global patterns inferred to exist from observations
and models of multiwavelength variability. The spectra of hot Jupiters
vary strongly over their surface as the effects of irradiation will
extremely vary over the surface; such effects are in principle
observable through high-precision photometric and spectroscopic
studies.

\subsection{Light Curves}\label{subsectionlightcurves}

\begin{figure}[H]
\centering
\includegraphics[width=12.2cm]{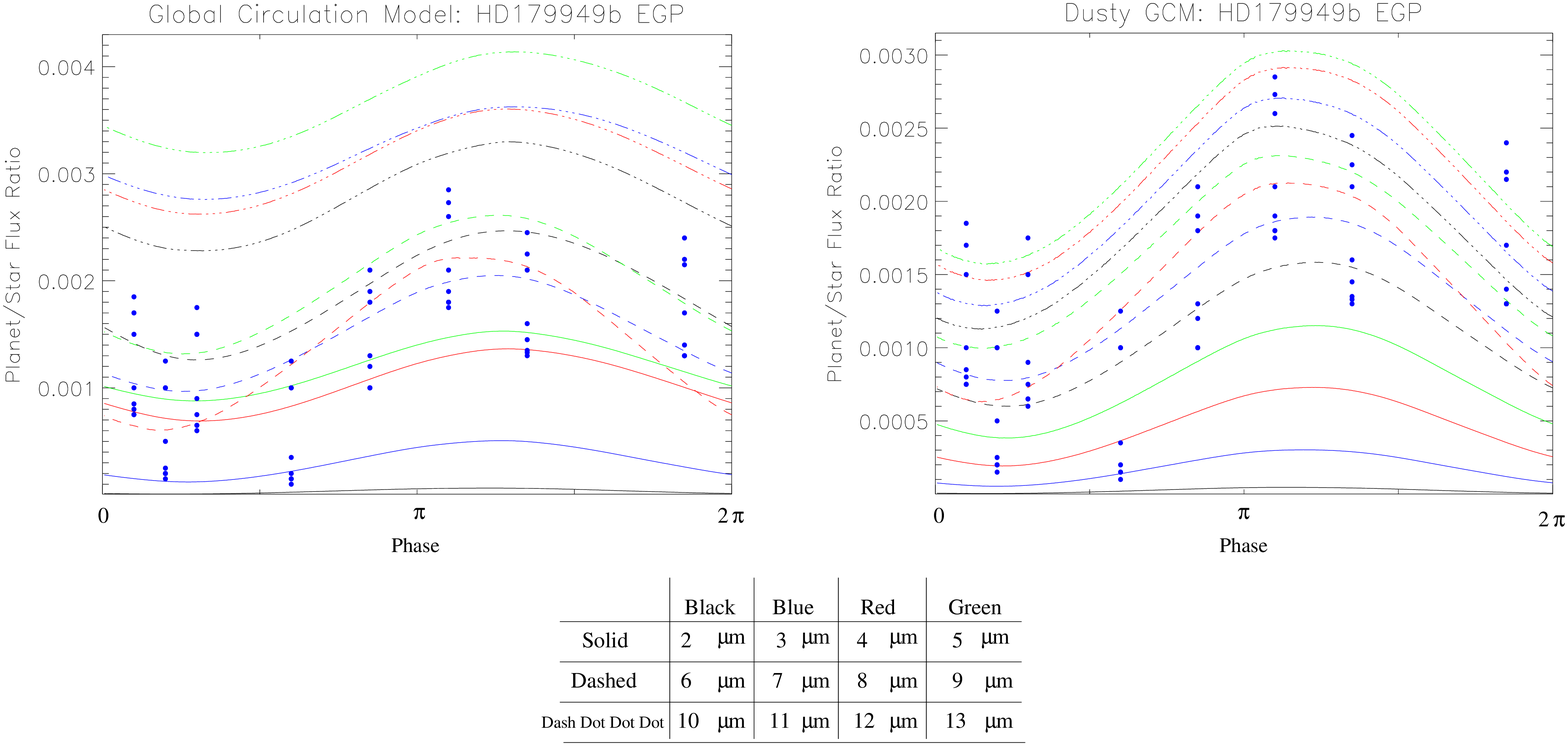}
\vspace{0.1cm}
\caption{\scriptsize Theoretical light curves in the 2.0-13.0 $\mu$m
  wavelength range for HD179949b. Left and right panels show
  cloud-free and dusty light curves, respectively. Within each panel,
  moving from top to bottom, the light curves are for wavelengths from
  13 $\mu$m to 2 $\mu$m. Blue dots correspond to the 8 $\mu$m data
  points by Burrows et al. (2008). Light curves reach their peak flux
  after the secondary eclipse. These phase offsets result directly
  from the atmospheric dynamics of the 1DRT
Two-Stream model.}
\label{fig.dyna8}
\end{figure}

In this subsection theoretical light curves as function of wavelength
for HD179949b have been calculated and confronted with actual
measurements by the Spitzer Space Telescope. The observational data
for the 8 $\mu$m light curve of HD179949b were taken from the Burrows
et. al (2008) work (Figure 7 there), just from them. That Figure 7
portrays a comparison between their theoretical models for various
combinations of parameters with the eight observational data points of
Cowan et al. (2007). The PHOENIX models simulate planetary transits,
for each time step, 360 in total, a summation (addition) over all
voxels over the planetary sphere gives the full intensity
integration. Then the intensity integration is used to generate the
PHOENIX light curves. The Global Circulation Model (hot Jupiter)
passing in front of its parent star is simulated with PHOENIX on a
257x257 pixels grid. The PHOENIX models assume circular orbits, which
is a good approximation for most exoplanets in tight orbits. Figure
\ref{fig.dyna8} presents theoretical light curves in the infrared
regime. It shows planet-star flux ratios, left and right panels
correspond to models for the cloud-free and dusty atmospheres,
respectively. Within each panel, moving from top to bottom, light
curves correspond to wavelengths from 13 $\mu$m to 2 $\mu$m.

Light curves in Figure \ref{fig.dyna8} reach their peak flux after the
secondary eclipse, a feature shared by all light curves. These phase
offsets result directly from the atmospheric dynamics in the 1DRT
Two-Stream model. According to Burrows et al. (2008), this may be a
feature of a hot Jupiter with stratosphere and/or hot upper
atmospheres. The shape and magnitude of the PHOENIX light curves
strongly depend on the Rayleigh scattering of light from the host
star, the presence of dust, the refraction of the stellar surface
brightness distribution, and the optical depth through the planet's
atmosphere, as determined by molecular opacity and clouds.

Then, the 8 $\mu$m cloud-free and dusty PHOENIX light curves were
compared to the published 8 $\mu$m data points by Burrows et
al. (2008). The theoretical and observational data are shown in Table
\ref{table.measurements}, first column corresponds to the orbital
phase; second and third columns to the PHOENIX planet-star Flux ratios
for the cloud-free ($R_{cf}$) and dusty ($R_d$) light curves,
respectively and fourth column ($R_{obs}$$\pm$${\sigma}$) to the
observational data and uncertainties of Burrows et al. (2008).

\begin{table}[H]
\scriptsize
\centering
\caption{Planet-star flux ratios and observational data for HD179949b ($R_{obs}$), the theoretical predictions correspond to cloud-free ($R_{cf}$) and dusty ($R_d$) models.}
\begin{tabular}{@{}lcccccccccccccc@{}}
\hline
Phase&Cloud Free (R$_{cf}$)&Dusty (R$_{d}$)&Data (R$_{obs}$$\pm$$\sigma$)\\
\hline
0.1&0.00065&0.00065&0.00125$\pm$0.0005\\
0.2&0.00061&0.00065&0.00075$\pm$0.00055\\
0.3&0.00068&0.00073&0.001175$\pm$0.000575\\
0.6&0.00122&0.00126&0.00065$\pm$0.00055\\
0.85&0.0018&0.00179&0.0015$\pm$0.0005\\
1.1&0.00221&0.00213&0.002375$\pm$0.000625\\
1.35&0.00213&0.00203&0.001875$\pm$0.000625\\
1.85&0.00104&0.00102&0.001875$\pm$0.000625\\
\hline
\end{tabular}
\label{table.measurements}
\end{table}

The chi-square statistics is defined by

\begin{equation}\label{eq.chisquares}
{\chi}{^2}  =  \sum\limits_{i=1}^{N} \frac{(R{_{cf}} - R{_{obs}})^2}{{\sigma_i}^2} 
\end{equation}

where the $\sigma_i$ is the uncertainty of each data point and $N$ is
the number of independent variables. Ideally, given the values of
$R_{cf}$ (or $R_{d}$) about their mean values $R_{obs}$, each term in
the sum will be of order unity. The degrees of freedom $\nu$ are equal
to 7, because there are $N$=8 data points and one adjustable
parameter. It is found that the reduced $\chi$$^2$ is equal to 1.24
for the cloud-free model and to 1.27 for the dusty model,
respectively. It would have had to have found the reduced $\chi$$^2$
in the vicinity of 1.0 to have been justified in suspecting the
consistency of the measurements. Discrepancies are discussed in the
discussion section.

\subsection{Reflection Spectra}\label{subsectiontransmissionspectra}

\begin{figure}[H]
\centering
\includegraphics[width=12cm]{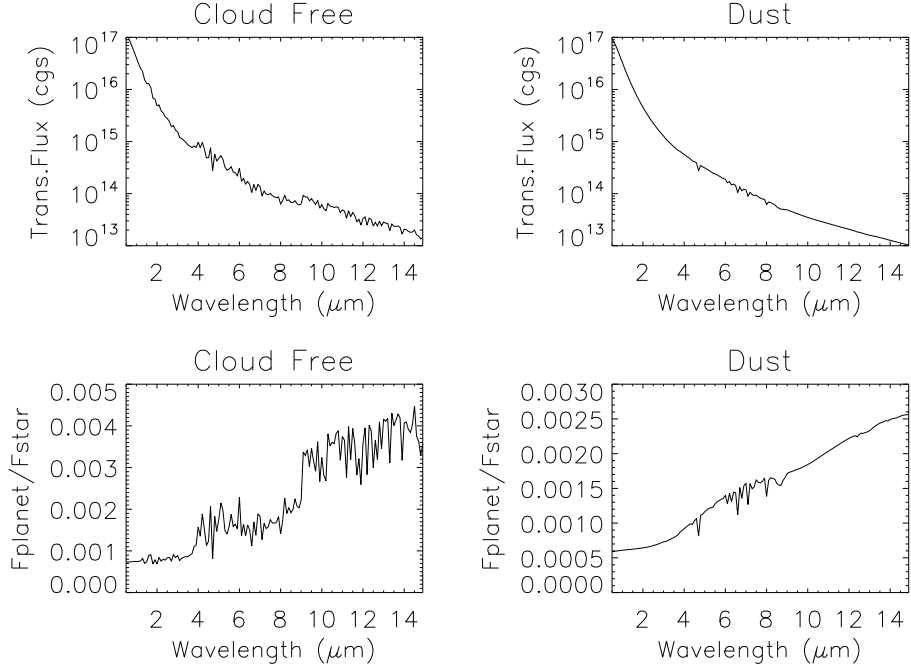}
\vspace{0.1cm}
\caption{\scriptsize 3D PHOENIX reflection spectra of the
  Planet HD179949b. Left frames correspond to
  cloud-free atmospheres whereas right frames to dusty
  atmospheres. Upper frames correspond to the stellar radiation
  through the planet's atmosphere, picking up some spectral features
  from the planet atmosphere and lower frames to the planet-star flux
  ratios. The cloud-free spectrum shows prominent H$_2$O molecular
  absorption features. Absorption features by dust opacities are
  smaller than the molecular absorption feautures.}
\label{fig.transmissionspectraNOhdust}
\end{figure}

The basic tenet of the PHOENIX reflection spectra method is that the
planet atmosphere absorption features are calculated as the stellar
flux passes through the planet's atmosphere above the limb. The
orbital phase for the PHOENIX reflection simulations is when the
planet is fully projected on the visible hemisphere of the star so
that the planet's projected transparent atmosphere takes out the
greatest area and limb darkening from the star is at its minimum. By
integrating the emergent intensities along an observer's line-of-sight
for the observer-limb-star orientation, the flux passing is calculated
through a section of a sphere encompassing the limb, in which case,
rays entering at different latitudes will pass through different
amounts of the planet's atmosphere. The theoretical reflection spectra
for the hot Jupiter HD179949b are shown in Figure
\ref{fig.transmissionspectraNOhdust}, there upper frames correspond to
the stellar radiation through the planet's atmosphere and bottom
panels to the planet-star flux ratios. Left frames correspond to
cloud-free atmospheres and right frames to dusty atmospheres,
respectively.

The PHOENIX atmospheres impose features on the spectra of its parent
star during transit; these features contain information about the
physical conditions and chemical composition of the atmospheres. In
the cloud-free spectrum, most prominent molecular absorption features
are seen from $\sim$4 $\mu$m to 14 $\mu$m. This spectrum shows
prominent H$_2$O molecular absorption features. In dusty models, the
most prominent absorption features are noted in the 4.5-9 $\mu$m
wavelength range, the 0.5-4.5 $\mu$m and 9-14 $\mu$m wavelength ranges
are featureless. Absorption features by dust are smaller than the
molecular absorption feautures. Many important optical/near-IR
molecular opacity sources (TiO and VO) have been completely removed by
the dust formation. It is also found that the apparent radius of the
planet in the optical regime is equal to 0.12 $R_{star}$, which is in
agreement with predictions in the literature (i.e. Wang $\&$ Ford
2011), the apparent radius is calculated following analysis of Seager
(2010).

\begin{figure}[H]
\centering
\includegraphics[width=10cm]{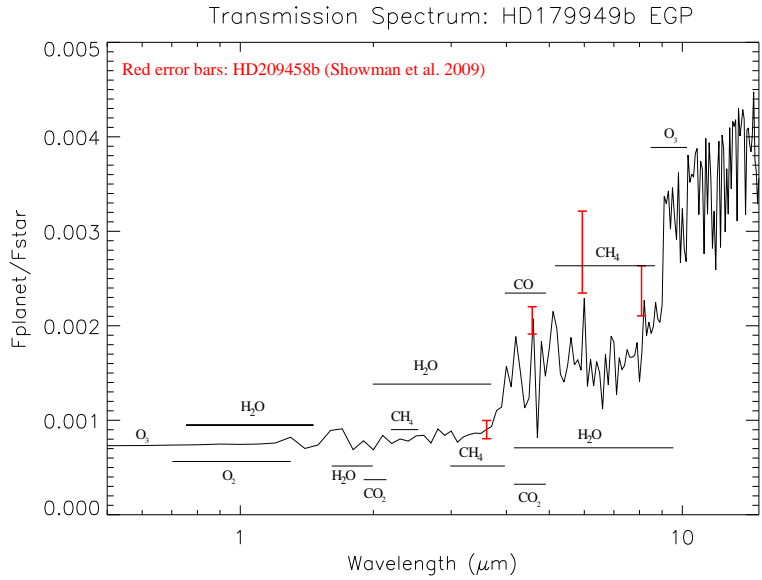}
\vspace{0.1cm}
\caption{\scriptsize Cloud-free reflection spectrum for the
  planet HD179949b. Absorption of visible and near IR radiation in the
  gaseous atmosphere are primarily due to H$_2$O, O$_3$, and CO$_2$,
  the main atmospheric gases absorbing/emitting in the IR are: CO$_2$,
  H$_2$O, O$_3$, CH$_4$, N$_2$O. In red error bars are those
  observational data for the hot Jupiter HD209458b. The PHOENIX model
  matches the secondary-eclipse depths at 3.6 $\mu$m, however, it
  underpredicts the eclipse depths at 4.5, 5.8 and 8.0 $\mu$m.}
\label{fig.spectrumnonirradiated}
\end{figure}

Then, the 3D PHOENIX reflection spectrum for HD179949b is confronted
with the Spitzer secondary-eclipse photometry of Knutson et al. (2008,
2009). The planet HD209458b was chosen to be confronted with the
PHOENIX model because it has similar orbital parameters to the hot
Jupiter HD179949b. The planet HD209458b is also interesting in terms
of reflection spectroscopy because of the presence of molecules such
as CH$_4$, CO and H$_2$O (Barman 2008; Swain et al. 2008). Table
\ref{table.measurementsspitzer} contains the observational planet-star
flux ratios for HD209458b and the PHOENIX predictions for HD179949b;
Figure \ref{fig.spectrumnonirradiated} shows the 3D PHOENIX cloud-free
reflection spectrum of HD179949b and in red error bars the
observational data of Showman et al. (2009) for the hot Jupiter
HD209458b.

\begin{table}[H]
\scriptsize
\centering
\caption{\scriptsize PHOENIX planet-star flux ratios for HD179949B and the observational data for HD209458B ($R_{obs}$).}
\begin{tabular}{|p{0.3\textwidth}|p{0.13\textwidth}|p{0.13\textwidth}|p{0.13\textwidth}|p{0.13\textwidth}|}
\hline
Planet-Star flux ratios (F$_P$/F$_S$)($\%$)&3.6 $\mu$m&4.5 $\mu$m&5.8 $\mu$m&8.0 $\mu$m\\
\hline
HD209458b ($R_{obs}$)&0.094$\pm$0.009&0.213$\pm$0.015&0.301$\pm$0.043&0.240$\pm$0.026\\
\hline
PHOENIX model for HD179949b&0.1044&0.14152&0.19024&0.174\\
\hline
\end{tabular}
\label{table.measurementsspitzer}
\end{table}

Figure \ref{fig.spectrumnonirradiated} shows important absorption
features, the most useful diagnostics are likely to be the
near-infrared bands of molecules, and the visible/near-IR resonance
lines of the alkali metals (Barman et al. 2005; Burrows et
al. 2008). In general, the cloud-free reflection spectrum in the mid
infrared is sensitive to molecular abundances. The strengths of
molecular lines provide diagnostics for the temperature, and the
vertical temperature structure. On the other hand, it is found that
the ${\chi}{^2}$ is equal to 37.12 and the reduced ${\chi}{^2}$ to
12.37 with 3 degrees of freedom. The PHOENIX spectrum in Figure
\ref{fig.spectrumnonirradiated} matches the secondary-eclipse depth at
3.6 $\mu$m, however, it underpredicts the eclipse depths at 4.5, 5.8
and 8.0 $\mu$m. When a spectrum exhibits greatly differing brightness
temperatures at different wavelengths (as in case of HD209458b), a
common explanation is that the different wavelengths sense different
pressure levels (because of the wavelength-dependent opacities)
(e.g. Cubillos et al. 2013). In this context, the fundamental
stumbling block to simultaneously explain the four Spitzer
secondary-eclipse depths is that the range of pressures that the
photons contribute to the 3.6, 4.5, 5.6, 8.0 $\mu$m wavelengths are
all very similar at least for the radiative-transfer PHOENIX model,
chemical composition, and opacities adopted for the model. Thus, it is
difficult to produce a high brightness temperature in some bands (such
as at 4.5 $\mu$m, 5.8 $\mu$m and 8$\mu$m) while maintaining low
brightness temperature in other bands (such as at 3.6 $\mu$m) as the
data apparently require. The temperature discrepancy problem is not
confined to the present study.

\section{Discussions and Conclusions}\label{sectionconclusiones}

In this work spectra and light curves of a detailed Global Circulation
Model (GCM) in the presence of a strong impinging radiation from its
parent star have been presented and compared to actual observations in
the infrared regime. The GCM uses radiative-equilibrium temperatures
as a function of longitude, latitude, and pressure; then the
Temperature-Pressure structures are employed as input structure files
to feed the 3D planetary atmosphere PHOENIX code. The incorporation of
a realistic 3D radiative transfer code represents the relevance of
this work in the development of hot jupiter atmospheres.

Radiative fluxes for the hot Jupiter HD179949b were calculated at the
substellar and anti-substellar points to look for differences. Models
on the day side show that the flux by the 3DRT PHOENIX code is larger
than the emission by the 1DRT Two-Stream model in the 5000 - 15000
{\AA} wavelength range. Nightside fluxes at the antisubstellar point
show deep absorption bands of H$_2$O, CO, and/or CH$_4$. Dusty models
on the day side show that the strong heating effects of dust opacities
prevent the formation of methane bands, and H$_2$O is dissociated
while producing a hotter water vapor opacity profile, which is much
weaker and more transparent to radiation. The dusty absorption
features are more prominent at the antisubstellar point than at the
stellar point. The dusty model generates a very smooth flux in the
optical as it is expected for an atmosphere dominated by grain
opacity.

Theoretical light curves as a function of wavelength for the hot
Jupiter HD179949b were calculated and compared to actual observational
data by the Spitzer Space Telescope. The PHOENIX simulations explain
the broad features of the observed 8.0 $\mu$m light curves (Cowan et
al. 2007, Burrows et al. 2008), including the fact that the
planet-star flux ratio peaks before the secondary eclipse. PHOENIX
radiative fluxes and phase-dependent light curves, reveal the
molecular and atomic compositions, atmospheric temperatures and the
degree to which the radiation on the day side is redistributed to the
night side.

The PHOENIX cloud free reflection spectrum was confronted with actual
data points from the Spitzer Telescope for the Jupiter HD209458b. The
theoretical predictions provides reasonable matches to the Spitzer
secondary-eclipse depth at 3.6 $\mu$m, however, it underpredicts the
eclipse depths at 4.5, 5.8 and 8.0 $\mu$m. These discrepancies result
from the chemical composition.

The majority of the problems that the results have matching the data
to the spectrum and light curves are not due to the PHOENIX
calculations but is rather due to the global circulation
calculations. Particularly, we are aware that better global
circulation models exist and have indeed been developed because of the
limitation of the newtonian cooling and its unability to fit the
observations. The GCM used in this work was created to simulate the
hot Jupiter HD209458b, which has been a prototypical exoplanet for
atmospheric thermal inversions (e.g. Showman et al. 2009), however
this assertion does not take into account recently obtained data or
newer data reduction techniques. Recently Diamond-Lowe et al. (2014)
revised the value of the Spitzer measurements and found a very good
agreement with the theory, they determined that it is unnecessary to
invoke a thermal inversion to explain the secondary-eclipse
depths. They concluded that there is no evidence for a thermal
inversion in the atmosphere of HD 209458b. This suggests the PHOENIX
models to select better circulation models for the hot Jupiter
HD179949b. Another possibility is that there is still physics that is
not included in the theoretical models which are causing the
discrepancies.

These assignments require a work collaboration with other groups to
produce one cohesive work. According to Diamond-Lowe et al. (2014),
their group is working on exploring the effects of a wide range of
planetary properties, including eccentricity, orbital distance,
rotation rate, mass, gravity, composition, metallicity, and stellar
flux on circulation models. Their work group could represent a
suitable collaboration to calculate theoretical radiative fluxes,
spectra and light curves of a variety of circulation models with the
radiative transfer PHOENIX code. We expect that the high spectral
resolution of JWST should enable meaningful inferences about
atmospheric structure and composition of hot Jupiters. The JWST will
be a powerful tool for exoplanet transit spectroscopy of a wide range
of planets, due to its broad wavelength coverage and high sensitivity
(Barstow et al. 2015).

Results presented in this study represent a first approach to a full
and detailed analysis of 3D spectra and light curves of hot
Jupiters. The relatively low resolution models presented here,
covering a broad wavelength range, place the overall features in
context, while models at medium or high resolution could be much
better suited to identify specific absorption features and probe
planetary atmospheres over a large pressure range.

Finally I conclude that there is a significant amount of work to be
done, it is especially crucial that better circulation models, high
resolution calculations, and full integrated radiative fluxes are
planned well in advance. Here we identify these specific areas for
necessary future work within the PHOENIX group.

\section{Acknowledgements}\label{sectionacknowledgements}

Author thanks the Graduiertenkolleg (GrK) 1351 Extrasolar Planets and
their Host Stars fellowship granted by the Deutsche
Forchunsgemeinschaft. He is greatly grateful to the Hamburger
Sternwarte of the University of Hamburg, Germany for a fruitful stay
where this work got started.

\end{document}